# Design and Development of a Java Parallel I/O Library

By

| | |
|---|---|
| Muhammad Sohaib Ayub | 2008-NUST-BIT-125 |
| Muhammad Adnan | 2008-NUST-BIT-140 |
| Muhammad Yasir Shafi | 2008-NUST-BEE-388 |

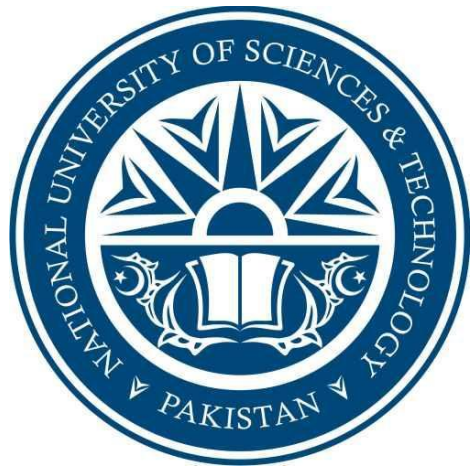

A project report submitted in partial fulfillment
of the requirement for the degree of
Bachelors of Information Technology

## Department of Computing

**School of Electrical Engineering and Computer Science
National University of Sciences & Technology
Islamabad, Pakistan
2012**

# CERTIFICATE

It is certified that the contents and form of thesis entitled **"Design and Development of a Java Parallel I/O Library"** submitted by **Muhammad Sohaib Ayub (2008-NUST-BIT-125), Muhammad Adnan (2008-NUST-BIT-140)** and **Muhammad Yasir Shafi (2008-NUST-BE-388)** have been found satisfactory for the requirement of the degree.

**Advisor:** \_\_\_\_\_\_\_\_\_\_\_\_\_\_\_\_\_\_\_\_\_\_\_\_\_\_\_\_\_

**(Syed. Akbar Mehdi)**

**Co-Advisor:** \_\_\_\_\_\_\_\_\_\_\_\_\_\_\_\_\_\_\_\_\_\_\_\_\_\_\_\_\_

**(Dr. Anjum Naveed)**

# DEDICATION

To Allah the Almighty

&

To my Parents and Faculty

# ACKNOWLEDGEMENTS

We are deeply thankful to Dr.Aamir Shafi for his throughout the project. His guidance, support and motivation enabled us in achieving the objectives of the project. We are extremely thankful to Ammar Ahmad Awan for guiding us in crucial stages of the project and sparing his valuable time for us. He was always helpful for us whenever we stuck in some complex problem.

We are deeply thankful to our Advisor, Syed Akbar Mehdi and Co-Advisor Dr. Anjum Naveed for helping us to improve our project throughout the course in accomplishing my final project. Their valuable suggestions were extremely helpful for us to achieve real targets for this project.

# TABLE OF CONTENTS

# LIST OF FIGURES

# LIST OF TABLES

# List of Abbreviations & Important Terms

**MPI –** Message Passing Interface

**MPJ -** Java based implementation of MPI

**HPC**- High Performance Computing

**ROMIO**- Implementation of MPI

**MPICH-2** Implementation of the message passing interface (MPI-2.2)

**JGF-** Java Grande Forum

**OpenMP-**Thread based API which provides parallelization mechanisms on shared-memory multiprocessors

**Parallel HDF5-** An API to support parallel file access for HDF5 files in a message passing environment.

**JNI-** Java Native Interface

**Java NIO-** Java New IO, an API which provides non-blocking I/O routines

# ABSTRACT

Parallel I/O refers to the ability of scientific programs to concurrently read/write from/to a single file from multiple processes executing on distributed memory platforms like compute clusters. In the HPC world, I/O becomes a significant bottleneck for many real-world scientific applications. In the last two decades, there has been significant research in improving the performance of I/O operations in scientific computing for traditional languages including C, C++, and Fortran. As a result of this, several mature and high-performance libraries including ROMIO (implementation of MPI-IO), parallel HDF5, Parallel I/O (PIO), and parallel netCDF are available today that provide efficient I/O for scientific applications. However, there is very little research done to evaluate and improve I/O performance of Java-based HPC applications. The main hindrance in the development of efficient parallel I/O Java libraries is the lack of a standard API (something equivalent to MPI-IO). Some adhoc solutions have been developed and used in proprietary applications, but there is no general-purpose solution that can be used by performance hungry applications.

As part of this project, we plan to develop a Java-based parallel I/O API inspired by the MPI-IO bindings (MPI 2.0 standard document) for C, C++, and Fortran. Once the Java equivalent API of MPI-IO has been developed, we will develop a reference implementation on top of existing Java messaging libraries. Later, we will evaluate and compare performance of our reference Java Parallel I/O library with C/C++ counterparts using benchmarks and real-world applications.

*Chapter 1*

# INTRODUCTION

Java is now considered as a mainstream programming language because of some very attractive features like built-in support for multi-threading, portability, automatic garbage collection, thread-safety, compile time and run time security. Consequently, Java has also been adopted by the High Performance Computing (HPC) community to program parallel applications for distributed and shared memory platforms. In modern HPC applications, disk I/O remains a significant bottleneck. Unfortunately, very little research has been done to evaluate and improve I/O performance of Java-based HPC applications. The main hindrance in the development of efficient parallel I/O Java libraries is the lack of a standard MPI-IO like API.

## 1.1 High Performance Computing

High Performance Computing is a branch of computer science that concentrates on writing high performance software on parallel hardware. One of the main areas of this discipline is to develop parallel processing algorithms and software, which are programs that can be divided into little pieces so that each piece can be executed simultaneously by a separate processor. Complex scientific problems have been solved for many years through high performance computing and facilities have been set up to run large number of codes in an efficient manner. HPC is used to solve many complex problems like climate modeling, turbulence, protein folding, patterns and speech recognition etc.

The calculation and solving of complex scientific problems can also be done by serial computing as well but there are several problems are associated with serial computing. First, the no of processors that work concurrently will always remain single, as it is serial computing. Second, the number of transistors present in single processor chip can only increase to a certain value showing the limited functionality of serial computing. Thus the need for parallel computing aroused that is the need to perform many tasks concurrently. Through parallel computing, the real-world scientific problems are solved in a less time as more number of processors are working to solve the same problem.

The scientific problems are solved by combining computer clusters and making all the clusters as a part of a single computational machine. This machine is then able to perform complex computations and calculations. The speed of computations and calculations has increased manifold in the high performance computing domain. However, the parallel I/O speed has not increased in the same pace, which is a great bottleneck in performing parallel I/O operation. Thus, the gap and difference of performance needs to be fulfilled between computations and parallel I/O. Although computational speed will always be greater than parallel I/O speed still the gap can be minimized.

HPC applications are written by following the message passing standard called as MPI (message passing interface). MPI is a specification of a message passing library. It is not a library itself but defacto standard for writing HPC applications. The implementation of MPI-I/O or MPI-2 standard is ROMIO. MPICH-2 is an implementation of the message passing interface (MPI-2.2) and ROMIO is part of the MPICH-2. It also provides a tool for MPI implementation research and for developing new and better parallel programming environments. MPICH-2 replaced MPICH1. Open MPI is a message passing interface (MPI) library combining technologies and resources from several other projects (FT-MPI, LA-MPI and PACX-MPI). It is used by many supercomputers with roadrunner, which was the world's fastest supercomputer from June 2008 to November 2009.

## 1.2   Java Based HPC

Although MPI in C and Fortran did become very popular in the beginning but there are some shortcomings in C. The MPI in C does not provide portability and multi-threading.

Java was considered as a very strong candidate in JGF (Java grande forum) and APIs were proposed in significant amount for implementation for MPI in Java. Thus two APIs were made in the name of mpiJava and MPJ. This significant number of implementations has been made with the name of MPJ /Express, mpiJava and MPJ/Ibis. MPJ/Ibis implements MPJ API and uses pure java devices that make use of java.io and java.nio packages to implement non-blocking and blocking communications at device level.

There has been very little research in I/O capabilities of Java but still there is no standard API for Java. Java has many qualities which makes it very easily the far better language than others. Java supports multi-threading, compile-time and run-time secure environment. Java provides automatic garbage collection unlike C. Also Java has the feature of portability due to which it is the most widely used language in the internet-based applications and shows that Java can perform on any platform given. Thus Java can prove to be better performing language than the other predecessors and we will use this language in our project to make a parallel I/O library.

## 1.3 I/O Requirements in High Performance Computing

Parallel I/O points to the capability of the program to read/write the data from/to the file using several processes which are present in distributed momory platforms.

Significant amount of increase has been done in storage and the computational power of the processors has been increased. Also the enhancements have been done in CPU and communication performance of the parallel machines. However same enhancements have not been made in I/O performance. The amount of storage has increased but the increase in the individual disk performance has not been done at the same pace. The peak performance has reached up to hundreds of Tflops /second of speed but the speed of I/O is only 100 Mb/s or less. Thus the speed of performance is very high but the I/O speed is low and is the bottleneck in applications requiring parallel I/O. This is the reason why parallel I/O libraries have been made in C, C++ and Fortran.

In the last two decades, there has been considerable amount of research in performing parallel I/O operations in languages like C,C++ and Fortran. As a result of this, several high performance libraries including ROMIO (implementation of MPI-IO), parallel HDF5, Parallel I/O (PIO), and parallel net CDF are available today that provide efficient I/O for scientific applications. However, very little research has been done to evaluate I/O in Java based HPC applications and the reason behind this is that no standard API has been developed for performing parallel I/O in Java. There is existing standard API like MPI-I/O in C which is not yet present in Java. Some amount of research has been done but there is no defacto standard which would help to enhance the parallel I/O capabilities in Java. Thus as part of the project we will develop a MPI-I/O equivalent implementation in Java which will be the standard parallel I/O library in Java. Existing projects on the same line include mpiJava, MPJ/Ibis.

There are two types of memory platforms, shared and distributed memory platforms. In shared memory platforms [Fig 2], all the tasks access the same memory while in distributed memory platforms [Fig 1]; the memory is local to all the processes and is shared. The distributed platform performs parallel I/O but the memory platform exists in the same place and all the processes perform the operation independently. Real parallel I/O is done by the shared memory platform as the memory is shared by the processes and all the processes work concurrently.

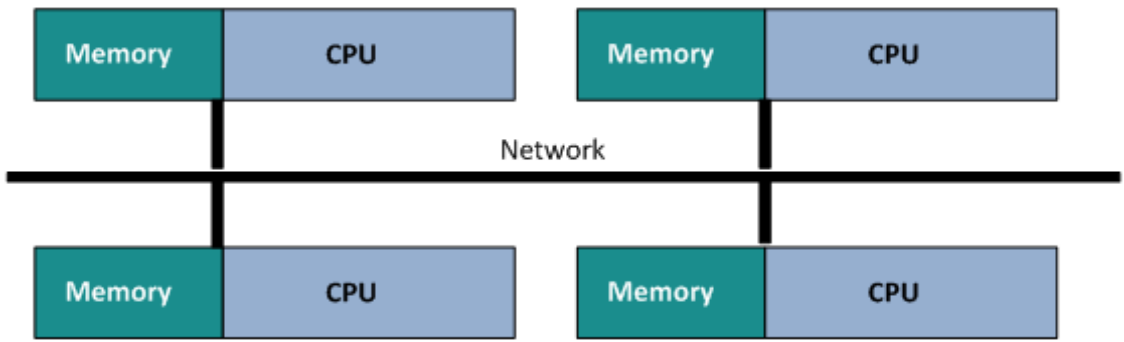

**Figure 1-1 Distributed Memory Architecture**

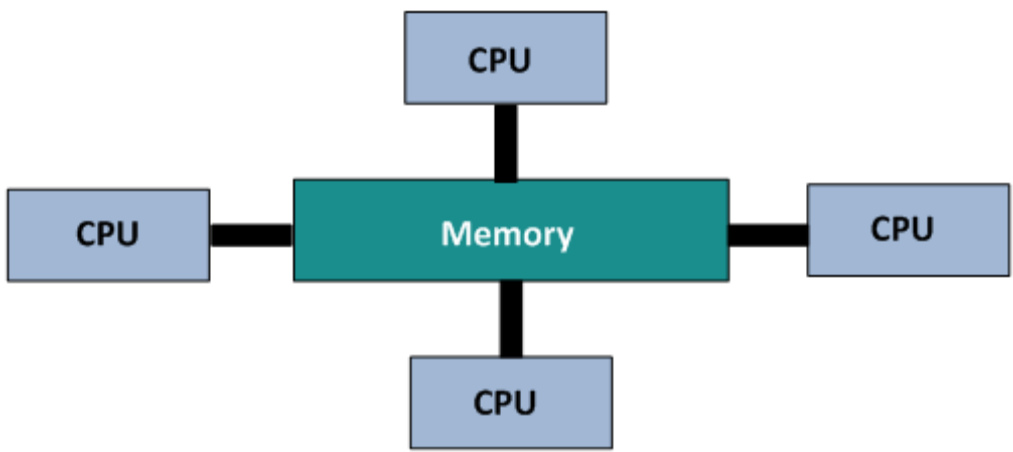

**Figure 1-2 Shared Memory Architecture**

### 1.3.1 Shared Memory Systems

Shared-memory systems are ones that have a common address space accessible to multiple nodes (processors). Shared memory locking mechanism is used to asynchronously read and write data from the memory. Such a model contains some inherent challenges such as race-conditions, deadlocks. Algorithms and techniques exist to avoid such challenges.

## 1.3.2 OpenMP

Open mp is an API based on threads and provides parallelization mechanisms on shared-memory multiprocessors. OpenMP defines compiler directives that specify regions of code that should be parallelized and define specific options for parallelization. Some pre-compiler tools also exist which can automatically convert serial programs into parallel programs by inserting compiler directives at appropriate places, making the parallelization of a program even easier.

OpenMP uses a thread-based fork-join model of parallel execution. The program runs a master thread serially until it reaches a directive to fork a team of threads that can be executed on different processors. Output from these threads can then be merged at the end of the parallel region, as master resumes its serial execution (till the next parallel directive). Fig [4] describes Fork/Join model adopted by OpenMP.

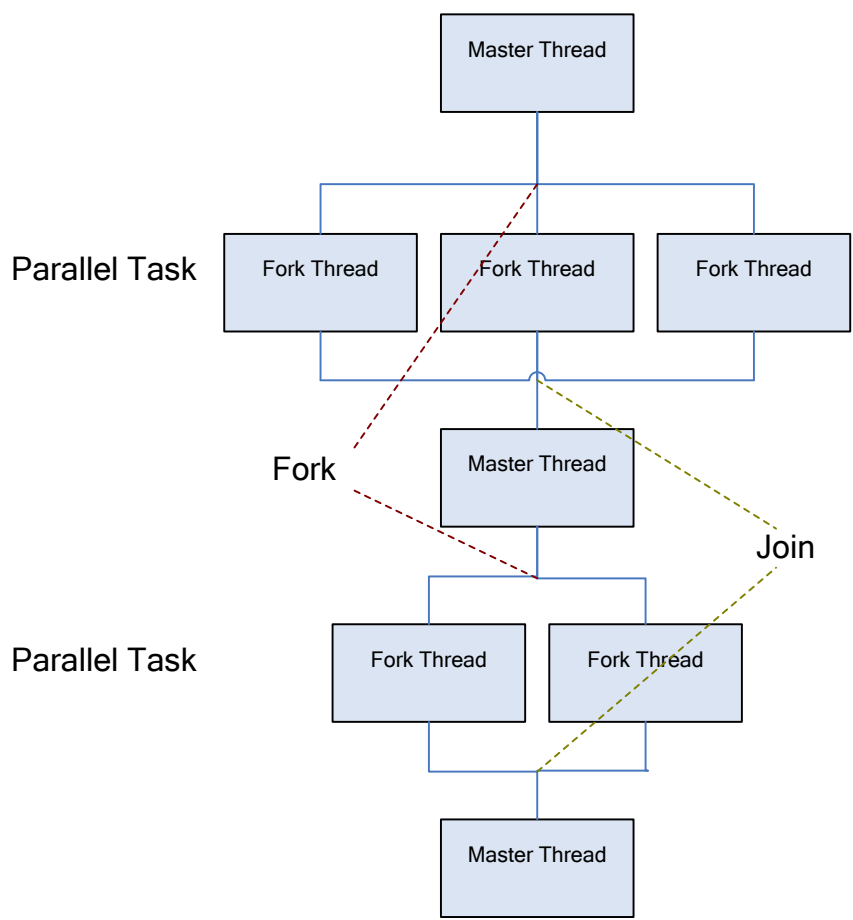

**Figure 1-3 Master thread forks a number of threads that execute code in parallel**

### 1.3.3 Parallel I/O Approach

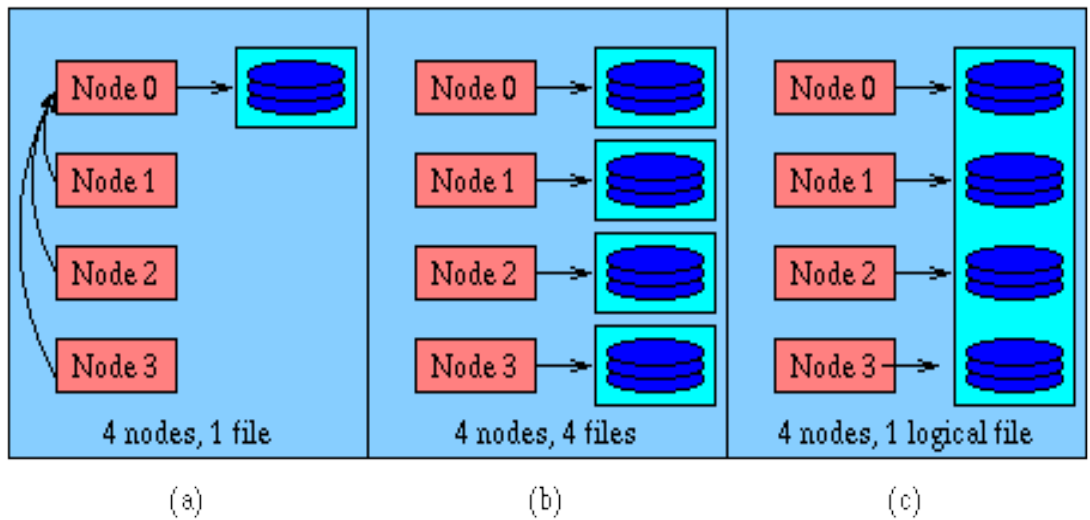

Figure 1-4 Parallel I/O Approach

Different approaches to performing parallel I/O have been shown In Fig 1-4. In a part, only one process or node1 is performing an I/O and the other three nodes are communicating with the node 0 which is performing the write operation on the file. This shows that the process is serial and also is slow as only one process is performing the operation. In part b four nodes or processes are performing their write operation concurrently and their operation is independent of the other processes. This is a parallel I/O operation and distributed memory architecture shown in Fig 6 is elaborated here. In part c, all of the four nodes are performing parallel I/O operation from the same storage.

## 1.4 I/O Requirements in Java

Parallel I/O research has been performed in C and language like C performs excellent operations in parallel I/O. In C, multidimensional arrays can be seen as one-dimensional array of the same length. C also allows casting of any type to the array of bytes. There is no need for synchronization while performing parallel I/O in C. Every process can access an independent regions of a shared random access file and perform the reads and writes in parallel to different areas of the file.

However due to some reasons like portability and multi-threading, other choices of languages have been considered which has lead to the choice of an object oriented language like Java. As written earlier, Java has many qualities which make it much more viable and useful for parallel I/O. The biggest feature of Java is its portability due to which this language has become the choice of many HPC applications.

Java I/O model is divided into two parts byte-oriented I/O and text-oriented I/O. Byte-oriented I/O relates to the data items of int, float, double etc and text-oriented is related to the characters and texts. Our main concern is the model for byte-oriented I/O. For byte-oriented I/O, the classes of input streams and output streams are used where stream is the ordered sequence of bytes.

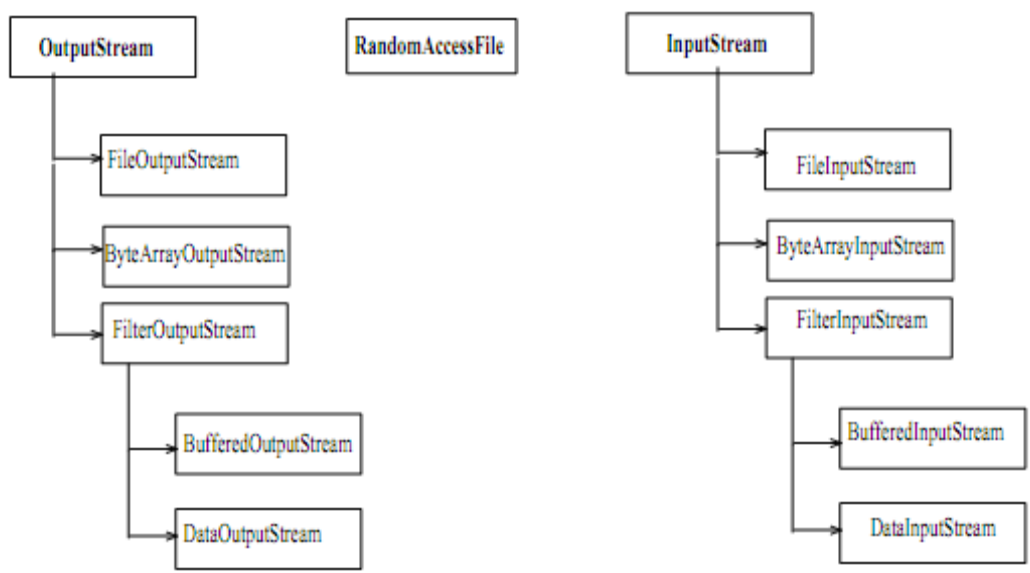

Figure 1-5 Java I/O Class Hierarchy

Input stream and output stream are classes and both consist of subclasses like file output stream, byte array output stream and filter output stream. Same for stands for the input stream hierarchy. Filter output stream consist of more subclasses. First one is the buffered output stream and second one is the data output stream. File output stream and file input stream are used to write and read to or from the file respectively. Byte array input and output stream is used to read or write from or to the byte array respectively. Filter streams give methods to chain streams together to build composite streams. Random access file block is not connected with any of the class hierarchies. Thus the data output stream can be chained to byte array output stream in order to reduce the number of calls to the file system

## 1.5 Problem statement

Therefore, from the discussion, it is clear that we need to design and develop a Java Parallel I/O Library. Since other native languages like C and Fortan have Parallel I/O libraries and we will compare the performance of our Java Parallel IO library with these existing libraries. Thus our problem statement is "Design and development of Java parallel I/O library and comparison of the performance with the C, C++ and Fortran counterparts through proprietary networks including Myrinet and Infiniband."

*Chapter 2*

# LITERATURE REVIEW

In the last two decades, there has been significant research in improving the performance of I/O operations in scientific computing for traditional languages including C, C++, and Fortran. As a result of this, several mature and high-performance libraries including ROMIO (implementation of MPI-IO), parallel HDF5, Parallel I/O (PIO), and parallel net-CDF are available today that provide efficient I/O for scientific applications. However, there is very little research done to evaluate and improve I/O performance of Java-based HPC applications. The main hindrance in the development of efficient parallel I/O Java libraries is the lack of a standard API (something equivalent to MPI-IO). Some ad-hoc solutions have been developed and used in proprietary applications, but there is no general-purpose solution that can be used by performance hungry applications.

## 2.1  MPI-IO

POSIX provides a highly portable file system but optimized parallel IO cannot be achieved with POSIX interface. Optimized and efficient I/O could be achieved if a standard exist which provides partitioning of the file among the processes and collective IO routines for memories and files. Efficiencies can be gained with support of asynchronous I/O accesses to the physical files and the storage devices. Instead of defining I/O access modes to express the common patterns for accessing a shared files (broadcast, reduction, scatter, gather), we chose another approach in which data partitioning is done in derived data types. This approach has advantage of flexibility and expressiveness as compared to previous approach of defining some routines for common patterns of broadcast, reduction, scatter and gather.

## 2.2  Parallel MPI IO in C, C++, Fortran

In last two decades, there has been significant research in improving the performance of I/O operations in scientific computing for traditional languages including C, C++, and Fortran. As a result of this, several mature and high-performance libraries including ROMIO (implementation of MPI-IO), parallel HDF5, Parallel I/O (PIO), and parallel netCDF are available today that provide efficient I/O for scientific applications.

### 2.2.1 ROMIO (Implementation of MPI-IO)

ROMIO is a high-performance, portable implementation of MPI-IO, and the I/O chapter in MPI-2. Version 1.2.5.1 of ROMIO (January 2003) is freely available.

ROMIO runs on at least the following machines: IBM SP; Intel Paragon; HP Exemplar; SGI Origin2000; Cray T3E; NEC SX-4; other symmetric multiprocessors from HP, SGI, DEC, Sun, and IBM; and networks of workstations (Sun, SGI, HP, IBM, DEC, Linux, and FreeBSD). Supported file systems are IBM PIOFS, Intel PFS, HP/Convex HFS, SGI XFS, NEC SFS, PVFS, NFS, and any UNIX file system (UFS).

#### 2.2.1.1 Features in ROMIO:

ROMIO is optimized for noncontiguous access patterns, which are common in parallel applications. It has an optimized implementation of collective I/O, an important optimization in parallel I/O. ROMIO 1.2.5.1 includes everything defined in the MPI-2 I/O chapter except support for file interoperability and user-defined error handlers for files. C, Fortran, and profiling interfaces are provided for all functions that have been implemented. It has implemented the subarray and distributed array datatype constructors from the MPI-2 miscellaneous chapter, which facilitate I/O involving arrays. It has also implemented the info functions from the MPI-2 misc. chapter, which allow users to pass hints to the implementation.

ROMIO is designed to be used with any MPI implementation. It is, in fact, included as part of several MPI implementations: Version, 1.2.5, is included in MPICH 1.2.5; an earlier version is included in LAM, HP MPI, SGI MPI, and NEC MPI. Version 1.2.5.1 is mainly a bug fix release.

ROMIO is freely available and is distributed in the form of source code.

### 2.2.2 Parallel HDF5

An API to support parallel file access for HDF5 files in a message passing environment. It provides fast parallel I/O to large datasets through standard parallel I/O interface. Processes are required to do collective API calls only when structural changes are needed for the HDF5 file.

Each process may do independent I/O requests to different datasets in the same or different HDF5 files. It supports collective I/O requests for datasets (to be included in next version). Minimize deviation from HDF5 interface.

#### 2.2.2.1 Features Supported in Parallel HDF5

It supports fixed dimension sized datasets, Extendible dimension sized datasets, Chunked storage datasets. Compression support: read only, no write. Data types: Integer, Float, String classes. Variable sized type support: read only, no write. I/O mode Independent read or writes. Collective read or write (collective as defined in MPI).

Limits Chunked storage (including extendible dimension sized datasets) does not support writing to overlapping chunks. That is process m and process n do not writing to the same chunk at the same time. If that happens, the result is undetermined. No write for compressed datasets. No write for variable length data types.

### 2.2.3 Parallel IO

PIO is an intermediate software layer the allows you to write data in either binary or netcdf format using either serial writes or parallel libraries such as MPI-IO and PnetCDF through a single interface.PIO is written in Fortran90.

The Parallel I/O (PIO) library was developed over several years to improve the ability of component models of the Community Climate System Model (CCSM) to perform I/O. However we believe that the interface is sufficiently general to be useful to a broader spectrum of applications. It currently supports netcdf, pnetcdf and MPI-IO.

PIO calls are collective, an MPI communicator is set in a call to PIO_init and all tasks associated with that communicator must participate in all subsequent calls to PIO. An application can make multiple calls to PIO in order to support multiple MPI communicators.

To use PIO your program should begin by calling the PIO_init function providing the MPI communicator and the rank within that communicator of the calling task. You should also provide the number of I/O tasks to be used, the stride or number of tasks between I/O tasks, and the number of MPI aggregators to be used. You may optionally also choose the base IO task; this task will be used for output of any non-decomposed data. This call initializes an IO system type structure that will be used in subsequent file and decomposition functions.

You can then open a file for reading or writing with a call to PIO_createfile or PIO_openfile. In this call you will specify the file type: pio_iotype_netcdf, pio_iotype_pnetcdf, or pio_iotype_binary, or the new netcdf4 types pio_iotype_netcdf4c, pio_iotype_netcdf4p; along with the file name and optionally the netcdf mode.

To read or write decomposed data you must first describe the mapping between the organization of data in the file and that in the application space. This is done in a call to PIO_initdecomp. In the simplest call to this function a one dimensional integer array is passed from each task the values in the array represent the 0 based offset from the beginning of the array on file.

### 2.2.4  Parallel NetCDF:

Parallel-NetCDF is a library providing high-performance I/O while still maintaining file-format compatibility with Unidata's NetCDF.

NetCDF gives scientific programmers a space-efficient and portable means for storing data. However, it does so in a serial manner, making it difficult to achieve high I/O performance. By making some small changes to the API specified by NetCDF, we can use MPI-IO and its collective operations.

Parallel-NetCDF makes use of several other technologies. ROMIO, an implementation of MPI-IO, provides optimized collective and noncontiguous operations. It also provides an abstract interface for a large number of parallel file systems. One of those file systems ROMIO supports is PVFS, a high performance parallel file system for Linux clusters.

## 2.3 Java based IO

There are very little Efforts done in evaluating and improving the I/O performance of HPC applications. The main hindrance in the development of efficient parallel I/O Java libraries is the lack of a standard API (something equivalent to MPI-IO). Some adhoc solutions have been developed and used in proprietary applications, but there is no general-purpose solution that can be used by performance hungry applications.

### 2.3.1 Evaluation of java's IO Capabilities

Java is becoming popular language for writing distributed applications because of its support for programming in distributed platform. It provides automatic garbage collection, run-time and Compile-time securities, portability, multithreading, support for persistent object and object migration, so there is a growing need for using java in HPC applications. Java should meet IO requirements of HPC to be successful in HPC domain.

This paper provides a detailed discussion and performance analysis of several approaches to parallel file IO available in java in two different parallel architecture and file system. Many scientific applications need to access large amount of data and IO often creates bottleneck in such applications. IO is divided into two parts in Java, one is byte-oriented which includes bytes, floats and integers and the other IO is text-based which includes character and text. Byte oriented IO in characterized by input streams and output streams. Java provides input stream and output stream classes for reading and writing.

Using raw byte arrays is an approach for performing parallel file IO in Java. If data is already in bytes form we can use java methods to read/write byte arrays however these java methods are only defined for byte-oriented data. If multiple threads of a parallel program need to write different parts of array to common file concurrently then we have to use threads in parallel and set offsets for random access to read/write objects in the shared file. This approach works correctly for both, when the new file is overwritten or new file is created because of the seek method semantics.

Mostly scientific applications operate on array of integers, floats and doubles instead of byte arrays but java provides no method for performing IO operations for arrays of integer, float or double so we explicitly convert byte-array of some other data type into an array of bytes and vice versa. We can write an array of integers by right shifting one byte at a time into a byte array and then writing the byte array while reading array of integers we can first read it into a byte array and then converting the byte array into integers. There is an issue of signed bit while converting integer array into bytes array or vice versa as java does not have unsigned data type so we take care of signed bit while conversion to/from array of bytes.

Java provides methods to read/write a single integer at a time in DataInput and DataOutput interface. RandomAccessFile class implements DataInput and DataOutput interfaces that is relatively easy to perform parallel IO using data streams. WriteInt is used to write single integer at a time.

Unbuffered data streams results in poorest performance because read write methods are called for each integer. RandomAccessFile class does not implement buffering and Filter Output/Input Streams only work with objects of InputStream and OutputStream .A random access file can be chained to a FileInputStream or

Design and Development of a Java Parallel I/O Library                                  Page 26

FileOutputStream object through file descriptor. FileInputStream or FileOutputStream object can be chained to BufferedInputStream or BufferedOutputStream objects which can be chained to DataInputStream and DataOutStream objects. Writing buffered data stream is unsafe for writing concurrently from multiple processes or threads to a common random access file because each thread maintains local buffer.

Using buffering with byte array stream is another approach to parallel file IO in Java. It allows buffering data input/output streams to chain it with underlying byte array stream .Read/write operations will be directed to byte streams rather than disk directly. After completion of writing toByteArray method is used to write data from byte array to shared file. Using byte arrays streams for reading operations are more complex. Each thread declares its own byte array and creates objects of ByteArrayInput and DataInputStream objects and seeks appropriate location in file. Each thread reads from the file into its byte array using read methods. Data is transformed from the byte array into the integer array using the read method of the data input stream class.

IO method are the only methods which provide reasonable IO performance .Real application need read write bulk of data which is mostly integer, float and double type. Read/write single integer or other data type results in poor performance. Java should provide Streams for read/write data and also provide its implementations to perform reasonable IO operation. Multidimensional arrays could be read/write by calling one dimensional array read/write methods. For high performance computing application developers would need a high-level parallel IO library (like MPI-IO) for java. Such library will benefit from the proposed methods for implementing reasonable parallel IO operations.

DataStream methods in Java provide poor performance even with careful selection of buffer size. To achieve reasonable performance, application is must use low-level IO methods to read/write array of bytes. Application must convert array of integer into array of byte or provide data stream methods in Java which could manipulate data type integer, float, double and other data types. This will be helpful for performing efficient parallel IO operations in java and will help in implementation of parallel IO library.

The main purpose of this study is to evaluate the support for efficient (and parallel) I/O in the Java programming language that can be used in high performance scientific applications. It identifies that I/O libraries provided by Java Development Kit (JDK) lack efficient support for array I/O operations of any data-type other than bytes. Most real-world scientific applications operate on data-types other than bytes as well. There are several ways to overcome this bottleneck—the study proposes the following approaches:

a) **Using Raw Byte Arrays:** Multiple threads of a parallel program can write to different parts of a byte array.

b) **Converting to/from an Array of Bytes:** Other array types can first be converted to/from byte arrays to perform efficient I/O operations. The conversion in this process becomes a major bottleneck.

c) **Using Data Streams:** Read/write a single integer at a time but it is extremely inefficient.

d) **Using Buffered Data Streams:** Utilize buffering for reading/writing to/from the files. This approach is not safe to use with multiple threads.

e) **Using Buffering with Byte Array Streams:** Buffer data to byte arrays and then perform I/O on disk.

*f)* **Bulk I/O:** Introduce bulk I/O operations for array data types other than bytes. Authors implement bulk I/O extensions on standard JVMs (using Java Native Interface (JNI)) and the Titanium language.

The JNI based bulk I/O extensions were originally presented in "Bulk File I/O Extensions to Java," but later became part of "Evaluation of Java's IO capabilities". These can be downloaded from [16]. The extensions include a Java class called BulkRandomAccessFiles. This class adds a few new methods, in addition to the java.io.RandomAccessFile class, for reading and writing either entire arrays of primitive data-types or a contiguous sub-sequence of elements in the array.

## 2.4 Existing Java parallel I/O projects

Java Sies, NetCDF Java library, Parallel Java (PJ) library, JExpand, Agent Team Work MPI-IO like library are the java based parallel IO libraries which are already created but most of them are not available for download or does not have parallel IO support or have some other issue so there are reasons for not recommending these libraries and there is need to develop a Java based parallel IO library which provides support for parallel IO.

### 2.4.1 Parallel Java library

PDF (Parallel Datastore System) is newly created system and allows read/write operations to/from a single logical storage files in parallel.PDF inspired by the likes of MPI-IO but is not compatible with any other popular APIs including MPI-IO.

Two group of classes are used, one for vectors and other for matrix.

PDF is like MPI-IO but not compatible with MPI-IO, parallel NetCDF or PIO. It makes use of PJ library for communication. Parallel Java is a pure Java, it does not support the native interconnect hardware like Myrinet and Infiniband.

### 2.4.2 Parallel file system

A parallel file system is one in which there are multiple servers as well as clients for a given file system, the equivalent of RAID across several file servers.

Parallel file systems are often optimized for high performance rather than general purpose use very large block sizes (=>64kB) and relatively slow metadata operations compared to reads and writes. They have special APIs for direct access Examples of parallel file systems: GPFS (IBM) Lustre (Cluster File Systems) PVFS2 (Clemson/ANL)

## 2.5 MPJ Express

Efforts to develop parallel applications in Java were proposed at the Java Grande Forum in 1998 not long after it was introduced as a language.

Earlier, developers had to choose between high performance and portability in parallel computing applications. PVM provided a highly portable approach to high performance computing; on the downside it was not comparable to the performance of the later MPI implementations in C and Fortran. MPI implementation in traditional C and Fortran have dominated as the most popular MPI implementation languages. These languages also lack modern language thread safety features, portability and object oriented approach.

MPJ Express provides Java bindings for MPI; it has built-in thread safety, a layered architecture and addresses the issues of high performance and portability by providing communication drivers using Java NIO (pure Java approach) and Myrinet.

Earlier efforts for building a Java messaging systems have typically followed either the JNI approach, or the pure Java approach. On commodity platforms advances in JVM technology now enables networking applications written in Java to rival their C counterparts. Improvements in specialized networking hardware has meant cutting down communication costs to a couple of microseconds [8].

Plans intending to increase MPJ Express's performance to make it comparable with C-based libraries such as MPICH2 and OpenMPI are underway; these include adding support for native device libraries and extending support for Infiniband.

MPJ Express is a Java based implementation of MPI having features of thread safety, layered architecture and addresses the issues of high performance and portability by providing communication drivers using Java NIO (pure Java) and Myrinet which can be swapped at runtime.

MPJ Express is a pure Java based messaging system that implements the MPI functionality using sockets. An intermediate buffering layer has been implemented to avoid the buffer copying overheads, suffered by some existing Java messaging systems. MPJ Express has a layered design that allows incremental development, and provides the capability to update and swap layers in or out as needed. This helps mitigate the contradictory requirements of end users because they can opt to use high-performance proprietary network devices or choose the pure Java devices that use sockets. Figure 8 explains current MPJ Express design and different levels of the software: the MPJ API, high level, base level, mpjdev and xdev.

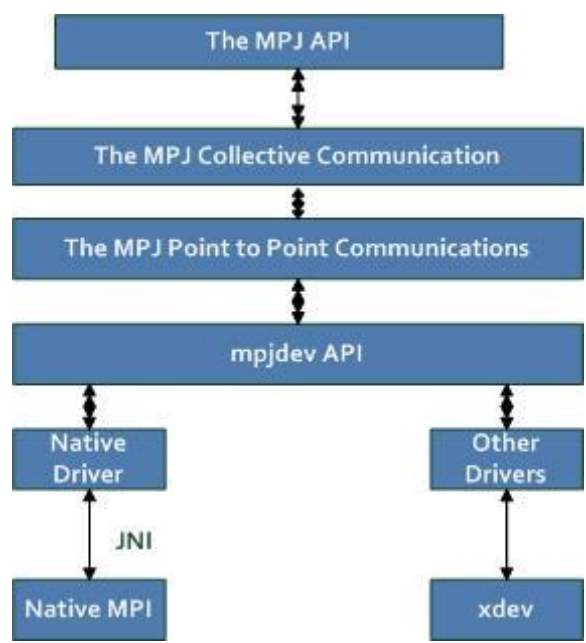

Figure 2-1 Layered architecture of MPJ Express

MPJ Express currently implements the mpiJava 1.2 API with a few changes. The top layer represents the exported API which defines approximately 125 functions. The next two layers implement collective communications and point to point communications. Collective communications are implemented using point to point communications.

*Chapter 3*

# DESIGN AND IMPLEMENTATION

We have developed a Java-based parallel I/O API and its reference implementation. This prototype implementation will become part of the MPJ Express (http://mpj-express.org) software, which is a Java MPI library developed at the HPC lab. There was no Java binding present for parallel I/O which could meet a standard for all performance hunger application developers to provide efficient I/O in Java based HPC. There exists MPI-IO standard library which provides routines and implementations which are performing efficient I/O.

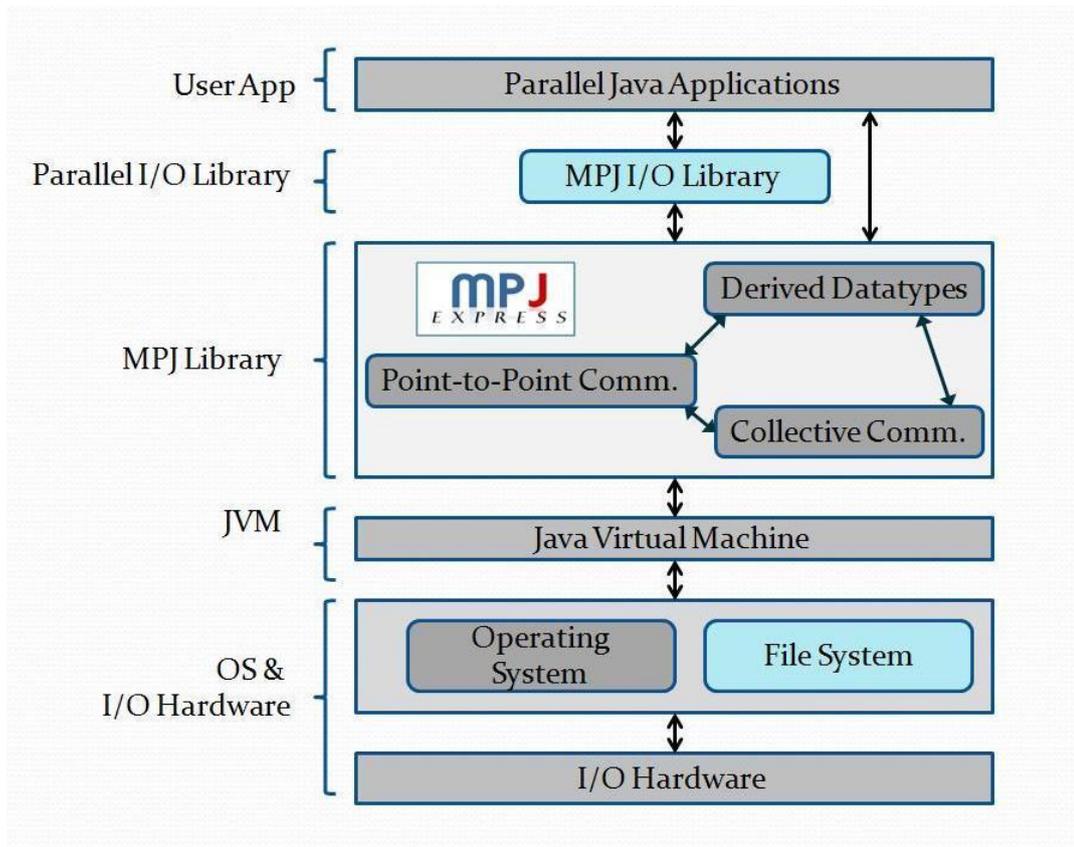

**Figure 3-1Architecture Diagram**

Java needs for development of distributed applications is growing because Java provides features of multi- threading, automatic garbage collection, compiles time and run time security and threads safety. These features provided by Java are the main reason for motivating programmers and developers to use it.

HPC involves lot of computations and lot of work has been done in C and Fortran. There are APIs in these languages which provide I/O bindings and they are working efficiently. MPI-IO is a standard C based API which provides C and Fortran routines for I/O operation.

## 3.1 Java NIO Based Parallel I/O

Java NIO provides buffer classes for primitive data types therefore we don't need explicit or implicit conversions for I/O operations. It provides buffer classes for Int, float, long, char, double, short and Byte Buffer. Buffer class has three main parameters which are position, limit and capacity. The position specifies the buffer index to/from we need I/O operation. Limit determines the number of items we want to read/write and the capacity is the total buffer capacity.

### 3.1.1  Memory Mapped File

Memory Mapped File is a magic provided by NIO and is helpful to create and edit large file size which are hard to bring in memory .It takes them as large arrays and moves required pages into the memory on demand. It allows programmers to directly access the contents from memory. It is achieved by mapping whole file or only a segment of file into memory.

### 3.1.2  Java NIO facilities

Java IO API provides Streams for I/O operations while NIO provides channels for IO. We are using file channel class for the I/O to/from files. It is fast access to the memory than Streams I/O but it could have problem of page faults while accessing a page from memory which is not already mapped. Reading and writing on memory mapped file is done by operating system, so even if your Java Program crash after putting content into memory it will make file until OS is fine.

## 3.2 Evaluation of NIO Approaches for Parallel Java I/O

We have evaluated 4 approaches to first write 1G data to a file and then read that from the from the file. The approaches tested are as follows

- Using FileChannel with view Buffer
- Using FileChannel in MappedMode
- Using RandomAccesFiles
- Using BulkRandomAccesFiles

### 3.2.1 BulkRandomAccessFiles

BulkRandomAccesFiles uses extension of Java and titanium language and it's not a part of standard JDK.It provides methods for reading and writing arrays of primitive data types but we are interested in only java based I/O so we will not entertain it.

### 3.2.2 RandomAccessFiles

RandomAccesFiles approach duplicates the functionality of Input Stream and Output Stream. It implements the DataStream interface and provides I/O methods for primitive data types only one element at a time which is an overhead because lot of switches are required to read write based on the data size.

### 3.2.3 FileChannel with viewBuffer

In our FileChannel with viewBuffer approach we are using asIntBuffer () for Int data-type in our discussion but the same holds true for all other primitive data-types. A view buffer is simply another buffer whose content is backed by the byte buffer. We exploit this functionality, in our proposed approaches, to perform memory operations on the view buffer and use the backing ByteBuffer object for I/O operations on a file using the FileChannel object.

### 3.2.4 FileChannel in MappedMode

In our technique of MappedMode the memory mapping is done and a portion of memory is brought into memory so we can create and edit large files. It gives illusion of file existence in memory and I/O can be performed from main memory which is highly beneficial to enhance our bandwidth for I/O operations.

## 3.3 Proposed Approaches for Performing Parallel Java I/O

We have proposed only first two approaches after our evaluation and we will be using these approaches for designing and implementing our Java based Parallel I/O API. We are using only these approaches because they give better bandwidth and persistent results.

We are inspired by MPI-IO and have shortlisted 19 out of 52 methods for our prototype implementation of our proposed Java Parallel IO API. We has implemented highlighted methods given below as our prototype of Java Parallel I/O API.

## 3.4 Proposed MPI-IO routines for Developing Parallel I/O Java API

**File Manipulation**

```
int MPI File open(MPI Comm comm, char *filename, int amode, MPI Info info,
            MPI File *fh)
int MPI File close(MPI File *fh)
```

**File views**

```
int MPI File set view(MPI File fh, MPI Offset disp, MPI Datatype etype,
               MPI Datatype filetype, char *datarep, MPI Info info)
```

**File consistency**

```
int MPI File set atomicity(MPI File fh, int flag)
int MPI File get atomicity(MPI File fh, int *flag)
int MPI File sync(MPI File fh)
```

**Data Access with Individual File Pointers**

```
int MPI File seek(MPI File fh, MPI Offset offset, int whence)
```

```
int MPI File get position(MPI File fh, MPI Offset *offset)

int MPI File get byte offset(MPI File fh, MPI Offset offset,
                             MPI Offset *disp)
```

### Individual file pointers – non collective – blocking

```
int MPI File read(MPI File fh, void *buf, int count, MPI Datatype datatype,
                  MPI Status *status)
int MPI File write(MPI File fh, void *buf, int count, MPI Datatype datatype,
                   MPI Status *status)
```

### Individual file pointers – non collective – non-blocking & split collective

```
int MPI File iread(MPI File fh, void *buf, int count, MPI Datatype datatype,
                   MPI Request *request)

int MPI File iwrite(MPI File fh, void *buf, int count,
                    MPI Datatype datatype, MPI Request *request)
```

### Individual file pointers – collective – blocking

```
int MPI File read all(MPI File fh, void *buf, int count,
                      MPI Datatype datatype, MPI Status *status)

int MPI File write all(MPI File fh, void *buf, int count,
                       MPI Datatype datatype, MPI Status *status)
```

### Individual file pointers – collective – non-blocking & split collective

```
int MPI File read all begin(MPI File fh, void *buf, int count,
                            MPI Datatype datatype)

int MPI File read all end(MPI File fh, void *buf, MPI Status *status)

int MPI File write all begin(MPI File fh, void *buf, int count,
                             MPI Datatype datatype)

int MPI File write all end(MPI File fh, void *buf, MPI Status *status)
```

**Table 3-1 Data Access Routines**

| positioning | synchronism | coordination | |
|---|---|---|---|
| | | *noncollective* | *collective* |
| *explicit offsets* | *blocking* | MPI_FILE_READ_AT<br>MPI_FILE_WRITE_AT | MPI_FILE_READ_AT_ALL<br>MPI_FILE_WRITE_AT_ALL |
| | *nonblocking & split collective* | MPI_FILE_IREAD_AT<br><br>MPI_FILE_IWRITE_AT | MPI_FILE_READ_AT_ALL_BEGIN<br>MPI_FILE_READ_AT_ALL_END<br>MPI_FILE_WRITE_AT_ALL_BEGIN<br>MPI_FILE_WRITE_AT_ALL_END |
| *individual file pointers* | *blocking* | MPI_FILE_READ<br>MPI_FILE_WRITE | MPI_FILE_READ_ALL<br>MPI_FILE_WRITE_ALL |
| | *nonblocking & split collective* | MPI_FILE_IREAD<br><br>MPI_FILE_IWRITE | MPI_FILE_READ_ALL_BEGIN<br>MPI_FILE_READ_ALL_END<br>MPI_FILE_WRITE_ALL_BEGIN<br>MPI_FILE_WRITE_ALL_END |
| *shared file pointer* | *blocking* | MPI_FILE_READ_SHARED<br>MPI_FILE_WRITE_SHARED | MPI_FILE_READ_ORDERED<br>MPI_FILE_WRITE_ORDERED |
| | *nonblocking & split collective* | MPI_FILE_IREAD_SHARED<br><br>MPI_FILE_IWRITE_SHARED | MPI_FILE_READ_ORDERED_BEGIN<br>MPI_FILE_READ_ORDERED_END<br>MPI_FILE_WRITE_ORDERED_BEGIN<br>MPI_FILE_WRITE_ORDERED_END |

## 3.5 Implementation of proposed routines

We have provided the MPJ-IO specs based on MPI-IO specs defined in MPI-2.2 standard document. Our prototype implementation details of the proposed methods are given here.

| **MPJ Classes** | **MPI Operations** |
|---|---|
| mpj.File class | MPI_File |
| mpj.Info class | MPI_Info |
| mpj.Offset class | MPI_Offset |

### 3.5.1 File Manipulation

All the collective operations are based on Communicator objects. There are two type of communicators; Intracomm and Intercomm. The MPI-IO operations are relevant to Intracomm class so we will translate the collective *MPI_FILE_OPEN* and *MPI_FILE_CLOSE* operation to methods of the Intracomm class. We note that the mpj.File class used in the method signatures is not to be confused with java.io.File class. In this document, whenever File is used, it means the mpj.File class and not java.io.File class

### 3.5.1.1 Opening a File

The Java binding of the MPI operation *MPI_FILE_OPEN*.

**File File.open(String filename, int amode, Info info)**

File open is a collective operation. We have used file channels which are present in Java NIO (Just like streams in Java I/O) for opening connections. Once we open a file, all process in the same communication group can operate this opened file. The process of only rank 0 opens the file and file handler is obtained from the Object if it is not already exist, then file object is created. The filename is name of the file and amode specifies the mode in which we are going to open this file, it could be read, write or both read and write mode. We have used info class which provides methods to set information and to get the information of the file which we have opened. The open method returns file object which could be further manipulated by other methods.

### 3.5.1.2 Closing a File

The Java binding of the MPI operation *MPI_FILE_CLOSE*.

**void File. close()**

File close is a collective call. This method is called from Intracom.Once a file is closed, all the process in the communication group couldn't access that file for further manipulations. You have to reopen that file in order to perform I/O operations on that file. We close a file using file handler or file object.

### 3.5.1.3 File Info

The Java binding of the MPI operation *MPI_FILE_SET_INFO*.

**void File.setInfo(Info info)**

The Java binding of the MPI operation *MPI_FILE_SET_INFO*.It is a collective routine.We use MPI setinfo method to set the information of the Info Object. We then use *MPI_FILE_GET_*INFO which returns the info of the object.

### 3.5.2 File Views

The Java binding of the MPI operation *MPI_FILE_SET_VIEW*.

**void File.setView(Offset disp, Datatype etype, Datatype filetype,**
**String datarep, Info info)**

The setView routine changes the process's view of the data in the file. The start of the file is set to disp. We have made setoffset method to set disp. We determine the type of data by etype. File type and etype are set to same so that auto conversion in primitive data types could be done without any costly operation of type conversion.

The Java binding of the MPI operation MPI_FILE_GET_VIEW. StringBuffer is used so that the datarep is passed-by reference and the changes in the body of the method reflect to the object itself.

**void File.getView(Offset disp, Datatype etype, Datatype filetype,**
**StringBuffer datarep)**

### 3.5.3 Consistency and Semantics

Consistency semantics define the outcome of multiple accesses to a single file. All file accesses in MPI are relative to a specific file handle created from a collective open. The MPI-IO semantics have scope of communicator group to open the file. MPI-IO guarantees the concurrent nonoverlapping writes correctly and changes are visible immediately to the writing process immediately.

Consistency is achieved by atomicity and synchronization. Java binding for the MPI operation *MPI_FILE_SET_ATOMICITY*.

**void File.setAtomicity (boolean flag)**

Java binding for the MPI operation MPI_FILE_GET_ATOMICITY returns the current consistency semantics for data access operations on the set of file

Design and Development of a Java Parallel I/O Library                                          Page 39

handles created by one collective open. If flag is true, the atomic mode is enabled and if the flag is false then monatomic mode is enabled. We have implemented set Atomicity and get atomicity methods to set atomic mode and to get current semantics of data access operations on the set of handlers created by one collective open. .

Java binding for the MPI operation *MPI_FILE_SYNC*.

**void File. sync ()**

Calling MPI_FILE_SYNC with fh causes all previous writes to fh by the calling process to be transferred to the storage device. If other processes have made updates to the storage device, then all such updates become visible to subsequent reads of fh by the calling process.MPI_FILE_SYNC may be necessary to ensure sequential consistency in certain cases MPI_FILE_SYNC is a collective operation. Calling MPI_FILE_SYNC with fh causes all previous writes to fh by the calling process. The user is responsible for ensuring that all nonblocking requests and split collective operations on fh have been completed before calling MPI_FILE_SYNC otherwise, the call to MPI_FILE_SYNC is erroneous.

### 3.5.4  Data Access

#### 3.5.4.1  Data Access Routines

There are three aspects to data access routines, positioning, synchronizing and coordination. Positioning is done by explicit offsets or implicit offsets. Coordintion is no-collective and collective. The following combinations of these data access routines include two types of pointers which are individual file pointers and shared file pointers.

#### 3.5.4.2  Data Access with Individual File Pointers

MPI maintains one individual file pointer per process per file handle. The current value of this pointer implicitly specifies the offset in the data access routines described in this section. These routines only use and update the individual file pointers maintained by MPI.

After an individual file pointer operation is initiated, the individual file pointer is updated to point to the next etype after the last one that will be accessed. The file pointer is updated relative to the current view of the file.

Java binding for the MPI operation *MPI_FILE_READ*.

**Status File.read(Object buf, int bufOffset, int count,
                    Datatype datatype)**

This routine provides blocking non-collective read operation. This method reads the file using individual file pointer.

Java binding for the MPI operation *MPI_FILE_READ_ALL*.

**Status File.readAll(Object buf, int bufOffset, int count,
                    Datatype datatype)**

This is the collective version of the blocking read operation using individual file pointer.

Java binding for the MPI operation *MPI_FILE_WRITE*.

**Status File.write(Object buf, int bufOffset, int count,
                    Datatype datatype)**

This is a noncollective blocking write operation using individual file pointer

Java binding for the MPI operation *MPI_FILE_WRITE_ALL*.

**Status File.writeAll(Object buf, int bufOffset, int count,
                    Datatype datatype)**

This is the collective version of blocking write operation using individual file pointer.

Java binding for the MPI operation *MPI_FILE_IREAD*.

**Request File.iread(Object buf, int bufOffset, int count,
                    Datatype datatype)**

This is a java binding for nonblocking noncollective read operation.

Java binding for the MPI operation *MPI_FILE_IWRITE*.

**Request File.iwrite(Object buf, int bufOffset, int count,
                                    Datatype datatype)**

   This is a collective routine for performing nonblocking write operation.

Java binding for the MPI operation *MPI_FILE_SEEK*.

**void File.seek(Offset offset, int whence)**

   Seek method updates the individual file pointer according to the whence. Whence could be MPI_SEEK_SET, MPI_SEEK_CUR or MPI_SEEK_END .MPI_SEEK_SET sets the individual file pointer to offset, MPI_SEEK_CUR sets the individual file pointer to current pointer plus offset and the MPI_SEEK_END sets the individual file pointer to the end of file plus offset.

Java binding for the MPI operation *MPI_FILE_GET_POSITION*.

**Offset File.getPosition()**

   This method returns the current position of individual file pointer in etype units relative to the current view. The offset can be used in a future call to MPI_FILE_SEEK using whence = MPI_SEEK_SET to return to the current position. To set the displacement to the current file pointer position, first convert offset into an absolute byte position using MPI_FILE_GET_BYTE_OFFSET, then call MPI_FILE_SET_VIEW with the resulting displacement.

Java binding for the MPI operation *MPI_FILE_GET_BYTE_OFFSET*.

**Offset File.getByteOffset(Offset offset)**

   It converts a view-relative offset into an absolute byte position. The absolute byte position of offset relative to the current view of fh is returned in disp.It returns the absolute byte file offset.

## 3.6 Test cases

### 3.6.1 Coll_test.java

The test uses collective read and write operation to write and then read file. 1KB data is first written and then read in the file. Buffer of 1KB is made and then this buffer is written and then read to/from the file.

### 3.6.2 Async_test.java

The test uses non-blocking read and write operation to write and then read file. 1KB data is first written and then read in the file. Buffer of 1KB is made and then this buffer is written and then read to/from the file.

### 3.6.3 Atomicity_test.java

This test uses simple blocking read and write operation with an addition of set_atomicity () and get_atomicity() methods. Set_atomicity() sets the atomic mode to true or false and get_atomicity() outputs the boolean number.

### 3.6.4 Misc_test.java

This test also uses simple blocking read and write operations along with the other method calls like getPosition() ,getByteOffset() and seek(). GetPosition() gives the position of the individual file pointer in the file and getByteOffset() converts offset into byte position. 1KB data is first written and then read in the file. Buffer of 1KB is made and then this buffer is written and then read to/from the file.

### 3.6.5 Perf.java

This test gives the performance evaluation in MB/s. First, the simple read and write operations are performed without sync() method call and performance is evaluated in MB/s. After this operation, the same performance evaluation is done with the sync() method call and the program outputs the numbers in MB/s.

Architecture Hierarchy of Implementation is given below.

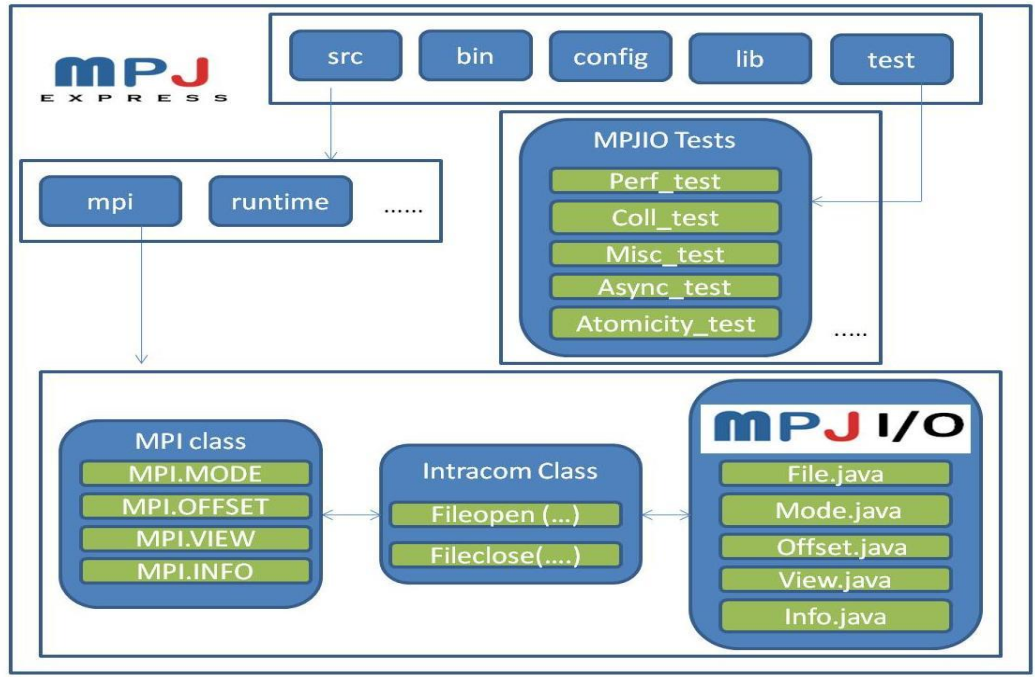

**Figure 3-2 Hierarchy of implementation**

*Chapter 4*

# EVALUATION AND DISCUSSION

## 4.1 Test Environment

Evaluation of parallel I/O approaches is performed on Barq and RCMS cluster. The specifications of these clusters are given below. These Clusters are configured at data centers at NUST H-12 campus.

### 4.1.1 Barq Cluster (barq.seecs.edu.pk)

The Barq cluster is composed of nine Intel Xeon Based Machines. The specification of Barq cluster is given below.

Table 4-1 Specification of Barq Cluster

| Cluster Name | Barq Cluster |
|---|---|
| Brand | Custom Built |
| Total Processors | 36 Intel Xeon |
| Total Nodes | Nine |
| Total Memory | 36 GB |
| Operating System | Open SuSE Linux 1.1 |
| Interconnects | Myrinet and Gigabit Ethernet |

**Specification of Barq Cluster**

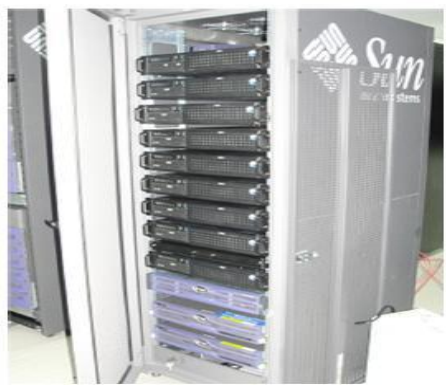

Figure 4-1 Barq Cluster Nodes

### 4.1.2 Afrit Cluster, RCMS (10.128.0.1)

The RCMS cluster is composed of 34 Intel Xenon based Machines and each of one is connected to NVidia Tesla S1070. All nodes are connected by 40Gbps interconnect Infiniband for internal communication. A high-performance and reliable SAN storage is linked by Servers, accessible by all computational nodes. The specifications of RCMS cluster are given below.

Table 4-2 Specification of RCMS Cluster

| Cluster Name | RCMS Cluster |
|---|---|
| Brand | HP ProLiant DL60se G6 Server/HP ProLiant DL380 G6 Server |
| Total Processor | 272 Intel Xeon |
| Total Nodes | 34 |
| Total Memory | 816 GB |
| Operating System | Redhat Enterprise Linux 5.5 |
| Interconnects | InfiniBand, Gigabit Ethernet |
| Storage | SAN storage 22TB raw capacity, SAN switches, host bus adapters, Fiber Channel Switch with RAID controller |
| Graphic Processing Unit | 32 x NVidia Tesla S1070 (each system contains 4 CPU's) |

**Specification of RCMS Cluster**

```
Node Names
U1/U2    C11    C22
C01      C12    C23
C02      C13    C24
C03      C14    C25
C04      C15    C26
C05      C16    C27
C06      C17    C28
C07      C18    C29
C08      C19    C30
C09      C20    C31
C10      C21    C32
```

Figure 4-2 Node structure of RCMS Cluster

## 4.2 Results of Parallel I/O Approaches

The results are for three approaches; "Using FileChannel with View Buffer", "Using FileChannel in Mapped Mode" and "Using BulkRandomAccessFiles".

### 4.2.1 Results of Parallel access to shared file residing on Local Disk

Figure 4-3 shows the results of tests executed on the shared memory machine, where Java threads were used for parallel access to the file residing on local disk. The read operation sustained a maximum bandwidth of approx. 10 GB/sec for file channel with view buffer approach. Exactly same trend and result was observed when the shared file was placed on NFS storage. Bulk random access file and file channel in mapped mode performed comparable for both the configurations, while file channel in mapped mode started to perform better for file on NFS storage as number of threads increased. This approach achieved a maximum bandwidth of 6 GB/seconds. The write operation for shared file on disk could only achieve a maximum bandwidth of 94 MB/sec. for all the three approaches. We experienced a sudden drop in performance for bulk random access file and file channel in mapped mode when number of threads increased from 4 to 8.

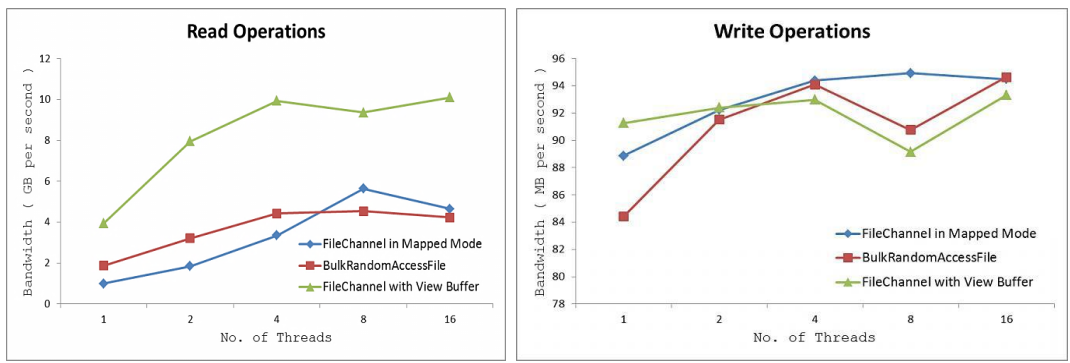

**Figure 4-3 Performance of Tests using Java threads for parallel access to a shared file on local disk**

### 4.2.2 Results of Parallel access to shared file residing on NFS

Figure 4-4 shows results when the shared file was moved to NFS storage instead of disk. We noticed that file channel in mapped mode performed inefficiently when file was moved to NFS storage. The reasons for this can be locking (mapping) mechanisms used by Java for memory-mapped regions of a file residing on NFS storage. Overall bandwidth increased significantly for file channel with view buffer and bulk random access file approaches, both achieved a maximum bandwidth of approx. 250 MB/sec. up from 94 MB/sec.

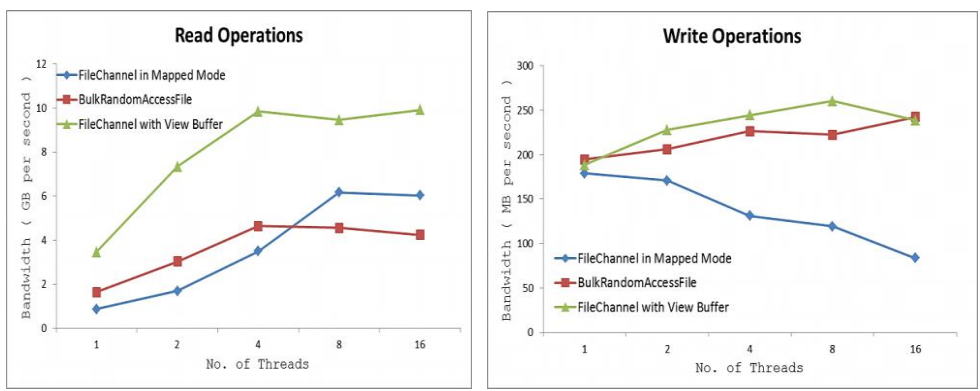

**Figure 4-4 Performance of Tests using Java threads for parallel access to a shared file residing on NFS storage attached to the Shared Memory Machine**

### 4.2.3 Results of parallel access to shared file residing on NFS storage of the Distributed Memory Machine

Results of tests executed on our distributed memory machine (cluster) are shown in Figure 4-5. We used MPJ Express processes instead of threads for tests on this machine. The read operation provided good speedups with increasing number of processes. File channel in mapped mode performed slower than other two approaches which achieved a maximum bandwidth of 40 GB/sec. for 24 processes. The write operation on the other hand saw a significant increase in performance, as number of processes increased from 16 to 24, for file channel in mapped mode approach and maximum bandwidth of 375 MB/sec. was achieved.

Performance of other two approaches was comparable and a maximum bandwidth of approximately 275 MB/sec was achieved.

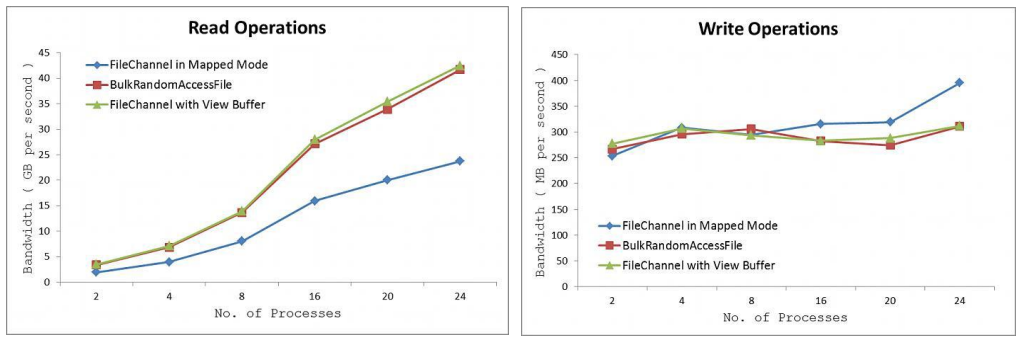

**Figure 4-5Performance of Tests using MPJ Express processes for parallel access to shared file residing on NFS storage of the Distributed Memory Machine**

The most stable performance across all configurations and tests was achieved by file channel with view buffers approach. Write performance increased significantly while read performance increased slightly, with increase in processes, for file channel in mapped mode on the distributed memory machine. Bulk random access file approach performed comparable to the file channel with view buffer approach.

## 4.3 Results of prototype implementation

Perf.java is the test case of our prototype implementation which gives the performance evaluation in MB/s. First, the simple read and write operations are performed without sync() method call and performance is evaluated in MB/s. After this operation, the same performance evaluation is done with the sync() method call and the program outputs the numbers in MB/s.

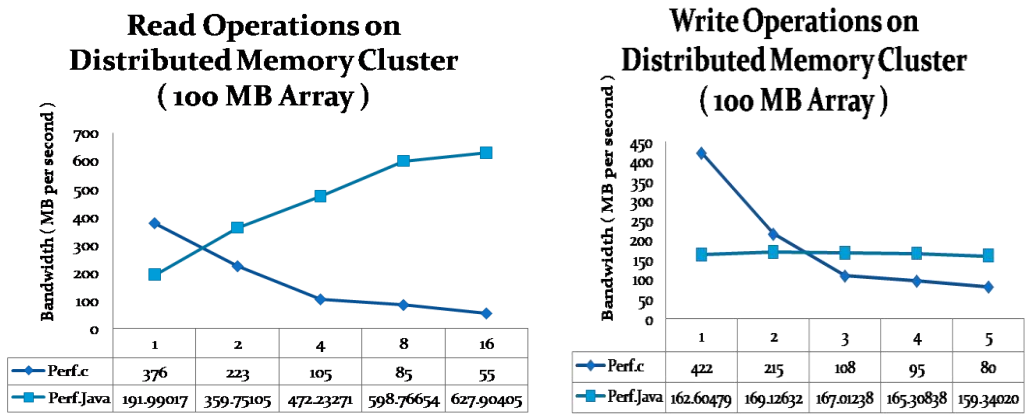

**Figure 4-6 Read and Write Test case results**

These are preliminary result. These may be different from results of complete implementation because we have not implemented consistency semantics for all file systems.

*Chapter 5*

# CONCLUSION AND FUTURE RECOMENDATION

Java has two I/O APIs, a legacy Java I/O API which was previously benchmarked and Java NIO API which has not been evaluated in the HPC context and is benchmarked in this paper. We observed that original Java I/O API provides poor file I/O performance, its extensions provided significant performance gains, and our proposed Java NIO approaches performed even better with increasing number of threads (and processes). In order to compete as a mainstream HPC language, Java based HPC libraries need to be equipped with efficient parallel I/O support. This can only be achieved if a standard MPI-IO style parallel file I/O API for Java be developed. Based on our performance evaluation, we can suggest that the design and implementation of a Java parallel I/O API shall be based on the Java NIO API as it natively provides the most efficient parallel file I/O methods. We are now focused on evaluation of Java NIO on different storage systems, as well as working on development of an MPI-IO style parallel file I/O API for Java.

Our work is an important step towards development of parallel I/O libraries for Java HPC. Our API is the first Java based API for parallel File I/O and The reference implementation of this API will also be the first ROMIO-like library for Java language. We have developed the first version of prototype implementation using MPJ Express we have implemented more than $1/3^{rd}$ part of complete reference implementation containing the essential functions.

The most stable performance across all configurations and tests was achieved by file channel with view buffers approach. Write performance increased significantly while read performance increased slightly, with increase in processes, for file channel in mapped mode on the distributed memory machine. Bulk random access file approach performed comparable to the file channel with view buffer approach.

Design and Development of a Java Parallel I/O Library    Page 51

We are interested in Java IO and our results show persistent, stable results with FileChannel with viewBuffer approach. We have used it for our prototype implementation and recommend this approach for performing the efficient I/O.

As a future work we will provide detailed implementation for all file systems, our current implementation is tested only on windows, Linux and NFS. Our current implementation could work on other file systems but unexpected could occur.

We have implemented prototype of our reference Java API and its complete implementation will be future work. The complete implementation will be tested for Windows, Linux, NFS and all file systems. There are two approaches for Implementation of complete Java parallel IO API.

- Using pure Java for all file systems
    - File system specific calls available is C but unexplored in Java
- JNI wrappers for ROMIO implementation

We will prove implementation of Info class to apply info hints for different file systems and others issues. These hints will be implemented in complete implementation.

File interoperability is not yet implemented even in ROMIO. Its implementation will be included in complete implementation of Java parallel I/O. File interoperability is the ability to read the actual bit information of the file which already has been written. Interoperability within a single MPI environment ensures that the data written by one MPI process can be read by any other MPI process, subject to the consistency constraints provided that it would have been possible to start the two processes simultaneously and have them reside in a single MPI_COMM_WORLD. Both of the processes must see the same data values at every absolute byte offset in the File for which data was written.

Our scope was MPJ Express which has not implemented required data types with holes for views that's why views are not implemented now and it will be in near future.

The test results in Figure [4-6] are preliminary result. These may be different from results of complete implementation because we have not implemented consistency semantics for all file systems.

*Chapter 6*

*Chapter 7*

# APPENDIX-A

## 7.1 Methods Implementation in Java based parallel I/OAPI

The Parallel I/O routines provided by MPI2.0 are listed in the following table. There are total 52 methods .we will implement (19 out of 52) methods highlighted in our prototype implementation of Java parallel I/O library (JPIO).

### 7.1.1 Data access routines

**Table 7-1 Data Access Routines**

| positioning | synchronism | coordination | |
| --- | --- | --- | --- |
| | | *noncollective* | *collective* |
| *explicit offsets* | *blocking* | MPI_FILE_READ_AT<br>MPI_FILE_WRITE_AT | MPI_FILE_READ_AT_ALL<br>MPI_FILE_WRITE_AT_ALL |
| | *nonblocking & split collective* | MPI_FILE_IREAD_AT<br><br>MPI_FILE_IWRITE_AT | MPI_FILE_READ_AT_ALL_BEGIN<br>MPI_FILE_READ_AT_ALL_END<br>MPI_FILE_WRITE_AT_ALL_BEGIN<br>MPI_FILE_WRITE_AT_ALL_END |
| *individual file pointers* | *blocking* | MPI_FILE_READ<br>MPI_FILE_WRITE | MPI_FILE_READ_ALL<br>MPI_FILE_WRITE_ALL |
| | *nonblocking & split collective* | MPI_FILE_IREAD<br><br>MPI_FILE_IWRITE | MPI_FILE_READ_ALL_BEGIN<br>MPI_FILE_READ_ALL_END<br>MPI_FILE_WRITE_ALL_BEGIN<br>MPI_FILE_WRITE_ALL_END |
| *shared file pointer* | *blocking* | MPI_FILE_READ_SHARED<br>MPI_FILE_WRITE_SHARED | MPI_FILE_READ_ORDERED<br>MPI_FILE_WRITE_ORDERED |
| | *nonblocking & split collective* | MPI_FILE_IREAD_SHARED<br><br>MPI_FILE_IWRITE_SHARED | MPI_FILE_READ_ORDERED_BEGIN<br>MPI_FILE_READ_ORDERED_END<br>MPI_FILE_WRITE_ORDERED_BEGIN<br>MPI_FILE_WRITE_ORDERED_END |

## 7.2 MPJ-IO: API Specification Version 0.1(MPI-2.2 Standard)

### 7.2.1 Introduction

There have been various additions to the original MPI-1.1 standard document. The final form that is available today is MPI-2.2 standard document.

The new MPI-3.0 is a work in progress and is still a draft version. We first need to address the changes from version 1.1 to version 2.2, as they are very important to extend the MPJ API and introduce parallel I/O functionality. The changes are listed below

### 7.2.1.1 Changes from MPI-1.1 to MPI-2.2 (relevant to MPI-IO interface)

1. Sub array data-type (chapter 4)
2. Distributed array data-type (chapter 4 – important for MPI-IO )
3. True extents of a data-type
4. Duplicate a data-type
5. Decoding a data-type
6. MPI IO Rank ( chapter 8 )
7. The MPI Info object ( chapter 9 )
8. MPI_COMM_JOIN ( chapter 10 )
9. Finalize method for I/O – ( chapter 12 )
10. MPI – I/O  ( chapter 13 )

Next, we provide the MPJ-IO specs based on MPI-IO specs defined in MPI-2.2 standard document. The sub-sections are laid out directly as in the standard document to allow cross-referencing. Since, the MPJ API specs define Java classes for opaque objects in the MPI standard; we have to define new classes that will reside in the mpj package.

| MPJ Classes | MPI Operations |
|---|---|
| mpj.File class | MPI_File |
| mpj.Info class | MPI_Info |
| mpj.Offset class | MPI_Offset |

### 7.2.1.2   Definitions
No special issues for the Java binding.

### 7.2.2   File Manipulation

All the collective operations are based on Communicator objects. There are two type of communicators; Intracomm and Intercomm. The MPI-IO operations are relevant to Intracomm class so we will translate the collective *MPI_FILE_OPEN* and *MPI_FILE_CLOSE* operation to methods of the Intracomm class. We note that the mpj.File class used in the method signatures is not to be confused with java.io.File class. In this document, whenever File is used, it means the mpj.File class and not java.io.File class

### 7.2.2.1 Opening a File

The Java binding of the MPI operation *MPI_FILE_OPEN*.

File File.open(String filename, int amode, Info info)

| | |
|---|---|
| filename | name of the file |
| amode | file access mode |
| info | the info object |
| returns: | the file object |

### 7.2.2.2 Closing a File

The Java binding of the MPI operation *MPI_FILE_CLOSE*.

void File.close()

### 7.2.2.3 Deleting a File

The Java binding of the MPI operation *MPI_FILE_DELETE*.

void File.delete(string filename, Info info)

| | |
|---|---|
| filename | the name of the file |
| info | the info object |

### 7.2.2.4 Resizing a File

The Java binding of the MPI operation *MPI_FILE_SET_SIZE*. We are using mpj.Offset object in the method signature for size parameter. The reason to use a special object rather than integer is that file offsets can easily be of sizes greater than $2^{32}$.

void File.setSize(Offset size)

size     size to truncate or expand file

### 7.2.2.5 Preallocating Space for a File

The Java binding of the MPI operation *MPI_FILE_PREALLOCATE*.

void File.preallocate(Offset size)

> size    size to pre allocate file

### 7.2.2.6 Querying the Size of File

The Java binding of the MPI operation *MPI_FILE_GET_SIZE*.

Offset File.getSize()

> returns:    the size of file in bytes

### 7.2.2.7 Querying File Parameters

The Java binding of the MPI operation *MPI_FILE_GET_GROUP*.

Group File.getGroup()

> returns:    the group that opened the file (mpj.Group)

The Java binding of the MPI operation *MPI_FILE_GET_AMODE*.

int File.getAmode()

> returns:    the file access mode of this file

### 7.2.2.8 File Info

The Java binding of the MPI operation *MPI_FILE_SET_INFO*.

void File.setInfo(Info info)

> info    the info object

The Java binding of the MPI operation *MPI_FILE_GET_INFO*.

Info File.getInfo()

> returns:    returns the info object

### 7.2.3 File Views

The Java binding of the MPI operation *MPI_FILE_SET_VIEW*.

void File.setView(Offset disp, Datatype etype, Datatype filetype,
                                String datarep, Info info)

> disp    displacement
> etype   elementary datatype objet (mpj.Datatype)

  filetype  filetype object (mpj.Datatype)
  datarep data representation
  info  the info object

The Java binding of the MPI operation *MPI_FILE_GET_VIEW*. StringBuffer is used so that the datarep is passed-by reference and the changes in the body of the method reflect to the object itself.

void File.getView(Offset disp, Datatype etype, Datatype filetype,
          StringBuffer datarep)

  disp  displacement
  etype  elementary datatype objet (mpj.Datatype)
  filetype  filetype object (mpj.Datatype)
  datarep data representation

### 7.2.4 Data Access

#### 7.2.4.1 Data Access Routines

There are three aspects to data access routines, positioning, synchronizing and coordination. Positioning is done by explicit offsets or implicit offsets.Coordintion is no-collective and collective. The following combinations of these data access routines include two types of pointers which are individual file pointers and shared file pointers.

#### 7.2.4.2 Data Access with Explicit Offsets

Java binding for the MPI operation *MPI_FILE_READ_AT*.

Status File.readAt(Offset offset, Object buf, int bufOffset, int count,
        Datatype datatype)
  offset the file offset
  buf  buffer object
  bufOffset the buffer offset
  count  number of elements in buffer
  datatype datatype of each buffer element
  returns: returns the status object

*MPI_FILE_READ_AT_ALL* is the collective version of the method above.

Status File.readAtAll(Offset offset, Object buf, int bufOffset,

                                int count, Datatype datatype)

    offset   the file offset
    buf          buffer object
    bufOffset    the buffer offset
    count        number of elements in buffer
    datatype     datatype of each buffer element
    returns:     returns the status object

Java binding for the MPI operation *MPI_FILE_WRITE_AT*.

Status File.writeAt(Offset offset, Object buf, int bufOffset,
                          int count, Datatype datatype)
    offset   the file offset
    buf          buffer object
    bufOffset    the buffer offset
    count        number of elements in buffer
    datatype     datatype of each buffer element
    returns:     returns the status object

Java binding for the MPI operation *MPI_FILE_WRITE_AT_ALL*. It is the collective version of the method above.

Status File.writeAtAll(Offset offset, Object buf, int bufOffset,
                            int count, Datatype datatype)
    offset   the file offset
    buf          buffer object
    bufOffset    the buffer offset
    count        number of elements in buffer
    datatype     datatype of each buffer element
    returns:     returns the status object

Java binding for the MPI operation *MPI_FILE_IREAD_AT*. This method is the non-blocking version of *MPI_FILE_READ_AT*

Request File.ireadAt(Offset offset, Object buf, int bufOffset,
                           int count, Datatype datatype)
    offset   the file offset
    buf          buffer object
    bufOffset    the buffer offset
    count        number of elements in buffer
    datatype     datatype of each buffer element

    returns:    returns the request object

Java binding for the MPI operation *MPI_FILE_IWRITE_AT*. This method is the non-blocking version of *MPI_FILE_WRITE_AT*

Request File.iwriteAt(Offset offset, Object buf, int bufOffset,
                        int count, Datatype datatype)

    offset    the file offset
    buf    buffer object
    bufOffset    the buffer offset
    count    number of elements in buffer
    datatype    datatype of each buffer element
    returns:    returns the request object

### 7.2.4.3 Data Access with Individual File Pointers

Java binding for the MPI operation *MPI_FILE_READ*.

Status File.read(Object buf, int bufOffset, int count,
                        Datatype datatype)

    buf    buffer object
    bufOffset    the buffer offset
    count    number of elements in buffer
    datatype    datatype of each buffer element
    returns:    returns the status object

Java binding for the MPI operation *MPI_FILE_READ_ALL*. This is the collective version of the above.

Status File.readAll(Object buf, int bufOffset, int count,
                        Datatype datatype)

    buf    buffer object
    bufOffset    the buffer offset
    count    number of elements in buffer
    datatype    datatype of each buffer element
    returns:    returns the status object

Java binding for the MPI operation *MPI_FILE_WRITE*.

Status File.write(Object buf, int bufOffset, int count,

                              Datatype datatype)

| | |
|---|---|
| buf | buffer object |
| bufOffset | the buffer offset |
| count | number of elements in buffer |
| datatype | datatype of each buffer element |
| returns: | returns the status object |

Java binding for the MPI operation *MPI_FILE_WRITE_ALL*. This is the collective version of the above.

Status File.writeAll(Object buf, int bufOffset, int count,
                              Datatype datatype)

| | |
|---|---|
| buf | buffer object |
| bufOffset | the buffer offset |
| count | number of elements in buffer |
| datatype | datatype of each buffer element |
| returns: | returns the status object |

Java binding for the MPI operation *MPI_FILE_IREAD*.

Request File.iread(Object buf, int bufOffset, int count,
                              Datatype datatype)

| | |
|---|---|
| buf | buffer object |
| bufOffset | the buffer offset |
| count | number of elements in buffer |
| datatype | datatype of each buffer element |
| returns: | returns the request object |

Java binding for the MPI operation *MPI_FILE_IWRITE*.

Request File.iwrite(Object buf, int bufOffset, int count,
                              Datatype datatype)

| | |
|---|---|
| buf | buffer object |
| bufOffset | the buffer offset |
| count | number of elements in buffer |
| datatype | datatype of each buffer element |
| returns: | returns the request object |

Java binding for the MPI operation *MPI_FILE_SEEK*.

void File.seek(Offset offset, int whence)

    offset   file offset
    whence      update mode

Java binding for the MPI operation *MPI_FILE_GET_POSITION*.

Offset File.getPosition()

    returns:      returns the file offset

Java binding for the MPI operation *MPI_FILE_GET_BYTE_OFFSET*.

Offset File.getByteOffset(Offset offset)

    offset   the view-relative offset
    returns:      returns the absolute byte file offset

### 7.2.4.4 Data Access with Shared File Pointers

Java binding for the MPI operation *MPI_FILE_READ_SHARED*.

Status File.readShared(Object buf, int bufOffset, int count,
                                      Datatype datatype)

    buf         buffer object
    bufOffset   the buffer offset
    count       number of elements in buffer
    datatype    datatype of each buffer element
    returns:    returns the status object

Java binding for the MPI operation *MPI_FILE_WRITE_SHARED*.

Status File.writeShared(Object buf, int bufOffset, int count,
                                      Datatype datatype)

    buf         buffer object
    bufOffset   the buffer offset
    count       number of elements in buffer
    datatype    datatype of each buffer element
    returns:    returns the status object

Java binding for the MPI operation *MPI_FILE_IREAD_SHARED*.

Request File.ireadShared(Object buf, int bufOffset, int count,
                                        Datatype datatype)

    buf          buffer object
    bufOffset    the buffer offset
    count        number of elements in buffer
    datatype     datatype of each buffer element
    returns:     returns the request object

Java binding for the MPI operation *MPI_FILE_IWRITE_SHARED*.

Request File.iwriteShared(Object buf, int bufOffset, int count,
                                        Datatype datatype)

    buf          buffer object
    bufOffset    the buffer offset
    count        number of elements in buffer
    datatype     datatype of each buffer element
    returns:     returns the request object

Java binding for the MPI operation *MPI_FILE_READ_ORDERED*.

Status File.readOrdered(Object buf, int bufOffset, int count,
                                        Datatype datatype)

    buf          buffer object
    bufOffset    the buffer offset
    count        number of elements in buffer
    datatype     datatype of each buffer element
    returns:     returns the status object

Java binding for the MPI operation *MPI_FILE_WRITE_ORDERED*.

Status File.writeOrdered(Object buf, int bufOffset, int count,
                                        Datatype datatype)

    buf          buffer object
    bufOffset    the buffer offset
    count        number of elements in buffer
    datatype     datatype of each buffer element
    returns:     returns the status object

Java binding for the MPI operation *MPI_FILE_SEEK_SHARED*.

void File.seekShared(Offset offset, int whence)

>   offset   file offset
>   whence      update mode

Java binding for the MPI operation *MPI_FILE_GET_POSITION_SHARED*.

Offset File.getPositionShared()

>   returns:   returns the file offset

### 7.2.4.5 Split Collective Data Access Routines

Java binding for the MPI operation *MPI_FILE_READ_AT_ALL_BEGIN*.

void File.readAtAllBegin(Offset offset, Object buf, int bufOffset,
                                        int count, Datatype datatype)

>   offset   the file offset
>   buf         the buffer object
>   bufOffset   the buffer offset
>   count      number of elements in buffer
>   datatype   datatype of each buffer element

Java binding for the MPI operation *MPI_FILE_READ_AT_ALL_END*.

Status File.readAtAllEnd(Object buf, int bufOffset)

>   buf         the buffer object
>   bufOffset   the buffer offset
>   returns:   returns the status object

Java binding for the MPI operation *MPI_FILE_WRITE_AT_ALL_BEGIN*.

void File.writeAtAllBegin(Offset offset, Object buf, int bufOffset,
                                        int count, Datatype datatype)

>   offset   the file offset
>   buf         the buffer object
>   bufOffset   the buffer offset
>   count      number of elements in buffer
>   datatype   datatype of each buffer element

Java binding for the MPI operation *MPI_FILE_WRITE_AT_ALL_END*.

Status File.writeAtAllEnd(Object buf, int bufOffset)

  buf   the buffer object
  bufOffset the buffer offset
  returns:  returns the status object

Java binding for the MPI operation *MPI_FILE_READ_ALL_BEGIN*.

void File.readAllBegin(Object buf, int bufOffset, int count,
              Datatype datatype)

  buf   the buffer object
  bufOffset the buffer offset
  count  number of elements in buffer
  datatype datatype of each buffer element

Java binding for the MPI operation *MPI_FILE_READ_ALL_END*.

Status File.readAllEnd(Object buf, int bufOffset)

  buf   the buffer object
  bufOffset the buffer offset
  returns:  returns the status object

Java binding for the MPI operation *MPI_FILE_WRITE_ALL_BEGIN*.

void File.writeAllBegin(Object buf, int bufOffset, int count,
              Datatype datatype)

  buf   the buffer object
  bufOffset the buffer offset
  count  number of elements in buffer
  datatype datatype of each buffer element

Java binding for the MPI operation *MPI_FILE_WRITE_ALL_END*.

Status File.writeAllEnd(Object buf, int bufOffset)

  buf   the buffer object
  bufOffset the buffer offset
  returns:  returns the status object

Java binding for the MPI operation *MPI_FILE_READ_ORDERED_BEGIN*.

void File.readOrderedBegin(Object buf, int bufOffset, int count,
                              Datatype datatype)

    buf        the buffer object
    bufOffset  the buffer offset
    count      number of elements in buffer
    datatype   datatype of each buffer element

Java binding for the MPI operation *MPI_FILE_READ_ORDERED_END*.

Status File.readOrderedEnd(Object buf, int bufOffset)

    buf        the buffer object
    bufOffset  the buffer offset
    returns:   returns the status object

Java binding for the MPI operation *MPI_FILE_WRITE_ORDERED_BEGIN*.

void File.writeOrderedBegin(Object buf, int bufOffset, int count,
                              Datatype datatype)

    buf        the buffer object
    bufOffset  the buffer offset
    count      number of elements in buffer
    datatype   datatype of each buffer element

Java binding for the MPI operation *MPI_FILE_WRITE_ORDERED_END*.

Status File.writeOrderedEnd(Object buf, int bufOffset)

    buf        the buffer object
    bufOffset  the buffer offset
    returns:   returns the status object

### 7.2.5 File Interoperability

File interoperability is the ability to read the actual bit information of the file which already has been written. Interoperability within a single MPI environment ensures that the data written by one MPI process can be read by any other MPIprocess, subject to the consistency constraints provided that it would have been possible to start the two processes simultaneously and have them reside
in a single MPI_COMM_WORLD. Both of the processes must see the same data values at every absolute byte offset in the File for which data was written.

#### 7.2.5.1 Datatypes for File Interoperability

Java binding for the MPI operation *MPI_FILE_GET_TYPE_EXTENT*.

int File.getTypeExtent(Datatype datatype)

    datatype    the datatype object
    returns:    returns the status object

#### 7.2.5.2 External Data Representation: "external32"

No special issues for the Java binding

#### 7.2.5.3 User-Defined Data Representations

Java binding for the MPI operation *MPI_REGISTER_DATAREP*.

void registerDatarep(String datarep, Object extraState)

    datarep    the datarep string
    extraState    extra state
    // callback discussion here. Signature incomplete

#### 7.2.5.4 Matching Data Representations

No special issues for the Java binding.

### 7.2.6 Consistency and Semantics

#### 7.2.6.1 File Consistency

Java binding for the MPI operation *MPI_FILE_SET_ATOMICITY*.

void File.setAtomicity(boolean flag)

    flag    the flag to set/unset the atomicity

Java binding for the MPI operation *MPI_FILE_GET_ATOMICITY*.

boolean File.getAtomicity()

returns: true if atomic mode, false if nonatomic mode

Java binding for the MPI operation *MPI_FILE_SYNC*.

void File.sync()

### 7.2.6.2 Random Access vs. Sequential Files

No special issues for the Java binding.

### 7.2.6.3 Progress

No special issues for the Java binding.

### 7.2.6.4 Collective File Operations

No special issues for the Java binding.

### 7.2.6.5 Type Matching

No special issues for the Java binding.

### 7.2.6.6 Miscellaneous Clarifications

No special issues for the Java binding.

### 7.2.6.7 MPI_Offset Type

MPI_Offset type is used instead of int datatype to represent the size of the largest file supported by MPI. The Java binding for this type is mpj.Offset.

### 7.2.6.8 Logical vs. Physical Layout

No special issues for the Java binding.

### 7.2.6.9 File Size

No special issues for the Java binding.

### 7.2.6.10 Examples

The examples in this section illustrate the application of the MPJ-IO consistency and semantics guarantees. These examples address

- Conflicting accesses on file handles obtained from a single collective open, and
- All accesses on file handles obtained from two separate collective opens.

Example -1: Sequential consistency by setting atomic mode

```
 1
 2    import mpj.*;
 3    class Example {
 5     public static void main (String args[]) {
 6          MPJ.Init(args);
 7          int myRank = MPJ.COMM_WORLD.rank();
 8          if (myRank == 0) {                    /* Process 0 */
 9              int a[] = new int[10];
10                 for (int i = 0; i < 10; i++) {
11                        a[i] = 5;
12                  }
13          File file = MPJ.COMM_WORLD.fileOpen("workfile",
14          MPJ.MODE_RDWR | MPJ.MODE_CREATE, MPJ.INFO_NULL);
15          file.setView(0, MPJ.INT, MPJ.INT, "native",
16           MPJ.INFO_NULL);
17            file.setAtomicity(true);
18            Status status = file.writeAt(0, a, 0, 10, MPJ.INT);
19            /* MPJ.COMM_WORLD.Barrier (); */
20         } else {  /* Process 1 */
21          int b[] = new int[10];
22           File file2 = MPJ.COMM_WORLD.fileOpen("workfile",
23            MPJ.MODE_RDWR | MPJ.MODE_CREATE, MPJ.INFO_NULL);
24        file2.setView (0, MPJ.INT, MPJ.INT, "native", MPJ.INFO_NULL);
25        file2.setAtomicity (true);
26            /* MPJ.COMM_WORLD.Barrier (); */
27            Status status = file2.readAt (0, b, 0, 10, MPJ.INT);
28       }
29        MPJ.Finalize();
30    }
31 }
```

Example 2: Alternate method of consistency by setting non-atomic mode and using barrier.

```
 1 import mpj.*;
 2
 3 class Example {
 4    public static void main (String args[]) {
 5
 6        MPJ.Init(args);
 7        int myRank = MPJ.COMM_WORLD.rank();
 8
 9        if (myRank == 0) {    /* Process 0 */
10
11            int a[] = new int[10];
12            for (int i = 0; i < 10; i++) {
13                a[i] = 5;
14            }
15            File file = MPJ.COMM_WORLD.fileOpen("workfile",
16                    MPJ.MODE_RDWR | MPJ.MODE_CREATE, MPJ.INFO_NULL);
17            file.setView(0, MPJ.INT, MPJ.INT, "native",
MPJ.INFO_NULL);

18            Status status = file.writeAt(0, a, 10, MPJ.INT);
19            file.sync();
20            MPJ.COMM_WORLD.Barrier ();
21            file.sync();
```

```
 22
 23          } else {      /* Process 1 */
 24
 25              int b[] = new int[10];
 26              File file2 = MPJ.COMM_WORLD.fileOpen("workfile",
 27                      MPJ.MODE_RDWR | MPJ.MODE_CREATE, MPJ.INFO_NULL);
 28                      file2.setView (0, MPJ.INT, MPJ.INT, "native",
MPJ.INFO_NULL);
 29             file2.setAtomicity (true);
 30              file2.sync ();
 31              MPJ.COMM_WORLD.Barrier ();
 32              file2.sync ();
 33              Status status = file2.readAt (0, b, 10, MPJ.INT);
 34          }
 35          MPJ.Finalize();
 36      }
 37 }
```

Example 3: erroneous attempt to achieve consistency by eliminating the apparently superfluous second "sync" call for each process in the example above.

```
 1 import mpj.*;
 2
 3 class Example {
 4     public static void main (String args[]) {
 5
 6         MPJ.Init(args);
 7         int myRank = MPJ.COMM_WORLD.rank();
 8
 9         if (myRank == 0) {     /* Process 0 */
 10
 11             int a[] = new int[10];
 12             for (int i = 0; i < 10; i++) {
 13                 a[i] = 5;
 14             }
 15             File file = MPJ.COMM_WORLD.fileOpen("workfile",
 16                     MPJ.MODE_RDWR | MPJ.MODE_CREATE, MPJ.INFO_NULL);
 17             file.setView(0, MPJ.INT, MPJ.INT, "native",
MPJ.INFO_NULL);
 18             Status status = file.writeAt(0, a, 10, MPJ.INT);
 19             file.sync();
 20             MPJ.COMM_WORLD.Barrier ();
 21
 22         } else {      /* Process 1 */
 23
 24             int b [] = new int[10];
 25             File file2 = MPJ.COMM_WORLD.fileOpen("workfile",
 26                     MPJ.MODE_RDWR | MPJ.MODE_CREATE, MPJ.INFO_NULL);
 27                      file2.setView (0, MPJ.INT, MPJ.INT, "native",
MPJ.INFO_NULL);
 28             MPJ.COMM_WORLD.Barrier ();
 29             file2.sync ();
 30             Status status = file2.readAt (0, b, 10, MPJ.INT);
 31         }
 32         MPJ.Finalize();
 33     }
 34 }
```

### 7.2.7 I/O Error Handling

No special issues for the Java binding.

### 7.2.8 I/O Error Classes

No special issues for the Java binding.

### 7.2.9 Examples

#### 7.2.9.1 Double Buffering with Split Collective I/O

Java version of the example in Section 13.9.1 of the MPI-2.2 standard document is below.

```
   /*
   ==============================================================
    *
    * Method: doubleBuffer
    *
    * Synopsis:
    * void doubleBuffer (MPJ.File file, MPJ.Datatype buftype, Int bufcount )
    *
    * Description:
    * Performs the steps to overlap computation with a collective writeby using
      a double-buffering technique.
    *
    * Parameters:
    * file       previously opened MPJ file handle
    * buftype    MPJ datatype for memory layout
    * (Assumes a compatible view has been set on file)
    *  bufcount    number of buftype elements to transfer
    *
   ==============================================================
   ======= */
```

```
1   void doubleBuffer (MPJ.File file, MPJ.Datatype buftype, int bufcount)
2   {
3        Status status; /* status for MPJ calls */
```

```
5            float buffer1 [] = new float [bufcount];
6            float buffer2 [] = new float [bufcount];
7
8            float computeBuf [];  /* destination buffer for computing */
9            float writeBuf [];    /* source buffer for writing */
10
11           int done; /* determines when to quit */
12           computeBuf = buffer1;  /* initially point to buffer1 */
13          writeBuf = buffer1;    /* initially point to buffer1 */
14
15   /* DOUBLE-BUFFER prolog:
16      * compute buffer1; then initiate writing buffer1 to disk
17    */
18     done = computeBuffer(computeBuf, bufcount);
19         file.writeAllBegin(writeBuf, 0, bufcount, buftype);
20
21        /* DOUBLE-BUFFER steady state:
22         * Overlap writing old results from buffer writeBuf
23         * with computing new results into buffer computeBuf.
24  * There is always one write-buffer and one compute-buffer in       use
25         * during steady state.
26         */

27         while (! done) {
28         toggleBuffer(computeBuf);
29         done = computeBuffer(computeBuf, bufcount);
30         status = file.writeAllEnd(writeBuf, 0);
31         toggleBuffer(writeBuf);
32             file.writeAllBegin(writeBuf, 0, bufcount, buftype);
33         }
34         /* DOUBLE-BUFFER epilog:
35         * wait for final write to complete.
36          */
37         status = file.writeAllEnd(writeBuf);
38     }
```

### 7.2.9.2  Subarray Filetype Constructor

No special issues for the Java bindings